\documentclass[12pt]{article}

\usepackage{fullpage,graphicx}
\usepackage{dcolumn}
\usepackage{bm}

\begin{document}

\title{Thermodynamic inadmissibility of the incompressible hydrodynamics \\
description of turbulent stratified fluid flows at low Mach number}
\author{R\'emi Tailleux \\
Department of Meteorology, \\ University of Reading, United Kingdom}
\date{11 November 2007}

\maketitle

\begin{abstract}
   The incompressible Navier-Stokes equations currently represent the primary
model for describing stratified turbulent fluid flows at low Mach number. The
validity of the incompressible assumption, however, has so far only been
rigorously established for adiabatic motions. Here, we show from first principl
es
that the use of available energetics and thermodynamics considerations applied
to a turbulent mixing event associated with stratified shear flow instability r
efutes
the widespread idea that the incompressible assumption is also valid when
diabatic irreversible effects are important. The main consequence is that
dynamics and thermodynamics are strongly coupled in stratified turbulence.
This departs strongly from the currently accepted wisdom, and calls for a
complete revisiting of the physical processes governing stratified turbulence
at low Mach numbers.

\end{abstract}

  The classical incompressible Navier-Stokes equations (INSE), often
in conjunction with the Boussinesq approximation \cite{Boussinesq1903},
currently represent
the primary model for the description and understanding of fluid flows
 at low Mach numbers $M_a=U_0/c_s$, where $U_0$ is the
characteristic fluid velocity, and $c_s$ the isentropic speed of sound.
However, despite its widespread use, the incompressibility assumption 
has been so far justified rigorously only in the context of adiabatic motions
\cite{Batchelor1967,Landau1987,Majda1984,Lions1998}, 
the generalization to flows with diabatic
and irreversible effects remaining an outstanding challenge
\cite{Ansumali2005}. Physically, the primary motivation for the
incompressibility assumption is arguably the simplification it brings 
about by decoupling the dynamics of the fluid from its thermodynamics.
While this appears of interest for the study of adiabatic motions,
for which thermodynamic effects are indeed of little importance,
the advantage of doing so otherwise is less clear, since it severely
complicates the understanding of two important fundamental issues, namely:
 1) whether the relaxation toward thermodynamic equilibrium plays any
role in stratified turbulent flows? 2) whether the work of 
expansion/contraction $B$ (see \ref{WBD_definitions} for a definition)
may sometimes contribute significantly to the production of mechanical
energy? Until now, these two issues have never been considered to be
relevant to understand the behavior of stratified turbulence, and are
therefore usually not discussed nor even mentioned in most current
reviews of stratified turbulence, e.g., \cite{Gregg1987,Fernando1991}.
Arguably, this is so because until now fluid dynamicists have lacked
a strong physical reason to question the validity of the INSE to
describe the behavior of turbulent fluid flows at low Mach numbers.
In this letter, we present what we believe is the first strong case
ever made against the validity of the INSE when irreversible diabatic
effects are present, by showing that the assumptions underlying
the INSE are inconsistent with exact energetics considerations based
on the full compressible Navier-Stokes equations (CNSE).

\par

%

 We start by pointing out the fact that, even though this is not universally
accepted, the INSE fail to describe the laminar evolution of a thermally
stratified fluid evolving under the action of molecular diffusion alone,
both qualitatively and quantitatively. The main reason that the incompressible
assumption does not make sense in that case is because the only kind of 
motion is due to the diabatic contraction/expansion of the fluid, which 
makes its velocity entirely divergent, with no solenoidal component.
 Moreover, it is also clear that the leading order energetics is inconsistent
with that of the INSE, as internal energy (IE) is in that case the primary
source of mechanical energy, with IE being first converted into kinetic
energy (KE), and then ultimately into gravitational potential energy (GPE). 
Physically, the ultimate cause for the motion here is the departure of
the fluid from thermodynamic equilibrium (TE thereafter), 
which molecular diffusion strives
to restore, in accord with the prediction of the second law of thermodynamics
stating that the entropy of an isolated system must increase until reaching
its state of maximum entropy. 
\par
If one agrees with this view, and hence with the primary role played by
the work of expansion/contraction $B$ in the laminar evolution of 
a stratified fluid, then it is natural to ask what could be so special
about turbulence that could relegate the role of $B$ to second order?
 Indeed, it is well accepted that entropy production (EP)
considerably increases
in a turbulent fluid, and it is even conjectured that EP could
possibly reach its maximum possible value 
permitted by circumstances \cite{Ozawa2001}, which is the well known
principle of maximum entropy production (MEP). An increase in entropy 
production, however, suggests that a turbulent fluid evolves more rapidly
toward its state of maximum entropy, i.e., toward thermodynamic equilibrium.
Such a conclusion seems unavoidable, unless one can come up with some
physical reason for believing that turbulence is able to generate some
kind of counter-processes able to halt entropy production at some
point before a state of maximum entropy is reached. We are skeptical
that such a reason can be physically derived, given that EP can only
stops when TE is reached, by which time the fluid would have become obviously
nonturbulent. This idea departs from the currently accepted wisdom, which
tends to consider that turbulent mixing homogenizes adiabatiatically 
conserved quantities. In the oceanic case, for instance, existing turbulent
mixing parameterization strive to homogenize potential temperature and
salinity, whereas TE considerations would suggest homogenizing in-situ
temperature and chemical potential \cite{Fofonoff1962} instead.
In any case, since the relaxation toward TE in the laminar case
occurs in conjunction with the ultimate conversion of IE into GPE via
the work of $B$, it seems legitimate to wonder whether in the turbulent
case a higher EP rate could possibly mean a faster IE/GPE conversion
via a correspondingly increased $B$? Physically, this is not inconsistent
with the idea that a turbulent nonlinear cascade toward smaller and smaller
scales should increase the curvature of the in-situ temperature field $T$,
which in turn should increase the magnitude of the local values of 
diabatic heating, with a corresponding increasing effect on the local
velocity divergence and ultimately on $B$ itself. In the classical view
based on the INSE, this scenario is eliminated by assumption, but as said
above it has never been discussed nor disproven. In fact, the main 
suprising result of this letter is precisely to suggest that this scenario
is precisely what characterizes the behaviour of stratified turbulent flows.
In order to establish our results, we take as our starting point the
fully compressible NSE written under the form:

\begin{equation}
     \rho \frac{D{\bf u}}{Dt} + \nabla P = - \rho g {\bf z} + 
     \nu \nabla^2 {\bf u}
    \label{momentum}
\end{equation}
\begin{equation}
     \frac{D\rho}{Dt} + \rho \nabla \cdot {\bf u} = 0
\end{equation}
\begin{equation}
        \frac{D\Sigma}{Dt} = \frac{\dot{Q}}{T}
\end{equation}
\begin{equation}
        I = I(\Sigma, \upsilon), \qquad
       T  = \frac{\partial I}{\partial \Sigma}, \qquad
       P = - \frac{\partial I}{\partial \upsilon}
\end{equation}
where ${\bf u}=(u,v,w)$ is the three-dimensional velocity field,
$P$ is the pressure, $\rho$ is the fluid density, $\upsilon=1/\rho$
is the specific volume, $g$ is the acceleration of gravity,
$\Sigma$ is the specific entropy, $\nu$ is the molecular viscosity,
$I$ is the specific internal energy, $T$ is the absolute temperature,
while the local diabatic heating is given by
\begin{equation}
   \dot{Q} = -\frac{1}{\rho} \nabla \cdot {\bf F}_q + \varepsilon
\end{equation}
where ${\bf F}_q = - \rho C_p \kappa \nabla T$ is the molecular 
diffusive flux of heat, with $C_p$ being the specific heat capacity
at constant pressure, and $\kappa$ the molecular heat diffusivity,
whereas $\varepsilon$ is the local dissipation rate of kinetic energy.
Note that in writing (\ref{momentum}), the Stokes' hypothesis of
vanishing bulk viscosity is made \cite{Graves1999} for simplicity,
but this could be relaxed by modifying appropriately the definition of the 
dissipation of kinetic energy in the following.

\par

  For simplicity, the fluid is assumed to be thermally and mechanically
isolated (except for the work of the surface atmospheric pressure 
$P_a$ against changes
in fluid volume). In that case, the
evolution equations for the kinetic energy (KE), gravitational potential
energy (GPE), and internal energy (IE) reduce to:
\begin{equation}
   \frac{d(KE)}{dt} = - W + B - D - P_a \frac{dV_{ol}}{dt}
   \label{KE_equation}
\end{equation}
\begin{equation}
   \frac{d(GPE)}{dt} = W
   \label{GPE_equation}
\end{equation}
\begin{equation}
   \frac{d(IE)}{dt} = D - B
   \label{IE_equation}
\end{equation}
where $V_{ol}$ is the fluid volume, 
while $KE$, $GPE$, and $IE$ refer to the volume-integrated
kinetic, gravitational potential, and internal energy respectively,
$$
  KE = \int_{V} \rho \frac{{\bf u}^2}{2}\, dV, \qquad
  GPE = \int_{V} \rho g z \, dV,
$$
\begin{equation}
   IE = \int_{V} \rho I \,dV
\end{equation}    
whereas the energy conversion terms $W$, $B$, and $D$ are given by:
$$
    W = \int_{V} \rho g w \, dV, \qquad
   B = \int_{V} P \frac{D\upsilon}{Dt} \rho\,dV,
$$
\begin{equation}
  D = \int_{V} \rho \varepsilon \, dV .
  \label{WBD_definitions}
\end{equation}
The focus here is on the energetics of an unstable
parallel stratified shear flow of the kind
that has been extensively studied by means of laboratory and
numerical experiments, e.g., \cite{Winters1995,Peltier2003}
and references therein. A typical
evolution involves three stages. First, a laminar evolution where
the shear flow and stratification are eroded through molecular 
viscosity and diffusion. Second, an unstable evolution associated
with the development of turbulent dissipative structures and intense
mixing. Last,
a return to laminar evolution once the conditions for instability
have been removed. Insight into the energetics of such a mixing event
can be obtained by integrating
Eqs (\ref{KE_equation}-\ref{IE_equation}) over the relevant time interval,
which yields the following budget equations for KE, GPE, and IE respectively:
\begin{equation}
    \Delta KE = -\overline{W} + \overline{B} - \overline{D}
    -P_a \Delta V_{ol}
\end{equation}
\begin{equation}
    \Delta GPE = \overline{W}
    \label{delta_gpe}
\end{equation}
\begin{equation}
     \Delta IE = \overline{D} - \overline{B}
\end{equation}
where the overbar and $\Delta (.)$ represent respectively a time integral
and difference over and between the endpoints of the time interval
considered. Thus, $\overline{W}$, $\overline{B}$, and $\overline{D}$ are
the total density flux, work of expansion/contraction, and KE dissipation
rate. It is customary to measure the ratio 
$\gamma_{mixing} = \Delta GPE/\overline{D}$, known as the ``mixing 
efficiency''. From (\ref{delta_gpe}),
one has $\overline{W}=\gamma_{mixing} \overline{D}$, which 
is a popular way to parameterize $\overline{W}$ \cite{Gregg1987}, 
for $\gamma_{mixing}$ appears to be reproducible observationally,
with a value of $0.2$ often cited and used \cite{Peltier2003}.
In the classical incompressible view of stratified turbulence, 
it is generally accepted to regard the works of expansion/contraction
due to $\overline{B}$ and $P_a \Delta V_{ol}$ as small or even negligible
compared to the density flux $\overline{W}$ or dissipation $\overline{D}$.
If so, the predictions of the incompressible model for the remaining
terms of the energy budget are:
\begin{equation}
    \frac{\Delta KE}{\overline{D}} \approx -(1+\gamma_{mixing}),
    \qquad \frac{\Delta IE}{\overline{D}} \approx 1,
    \label{boussinesq_prediction}
\end{equation}
\begin{equation}
    \frac{\Delta KE}{\Delta IE} \approx - (1+\gamma_{mixing})
   \label{boussinesq_prediction_2} .
\end{equation}
In the following, we show that the fully compressible NSE yield
distinct predictions for (\ref{boussinesq_prediction}), but similar 
for (\ref{boussinesq_prediction_2}). It follows that
accurate measurements of $\Delta KE$, $\Delta IE$, and $\overline{D}$
should in principle
allow Eq. (\ref{boussinesq_prediction}) to provide the basis
for an observational test of the incompressibility assumption,
but this remains to be carried out.

\par

 In order to make progress, we need an independent way to estimate the
magnitude of the work of expansion/contraction $\overline{B}$; as 
regards to the work of atmospheric pressure against volume changes,
we don't question the validity of assuming it to be negligible. In
this letter, additional information about $\overline{B}$
is derived by computing 
separate budgets for the available and un-available components of GPE
and IE, as defined by Lorenz \cite{Lorenz1955}. Such an idea was
originally used in the present context by \cite{Winters1995}, in the
framework of the Boussinesq approximation. Since we question the validity
of the latter, the work of \cite{Winters1995} is generalized to the
fully compressible NSE. To that end, let us recall that the concept
of APE is defined as the difference between the PE of the actual state
minus that of a reference state which is defined as the state minimizing
the total potential energy PE=GPE+IE in an adiabatic re-organization
of the fluid parcels \cite{Lorenz1955,Winters1995}. In mathematical terms,
the reference state is defined by an invertible mapping taking
a parcel at the position ${\bf x}$ to its position ${\bf x}_r=
{\bf x}_r({\bf x})$ in the reference state. By definition, such a mapping
conserves the mass and entropy of the fluid parcels, which implies:
\begin{equation}
     \Sigma ({\bf x}_r,t) = \Sigma ({\bf x},t) = \Sigma_r (z_r,t),
     \label{sigma_r}
\end{equation}
\begin{equation}
     \rho ({\bf x}_r,t) J_r  
  = \rho ({\bf x},t) = \rho_r (z_r,t) ,
   \label{rho_r}
\end{equation}
where $J_r = \partial ({\bf x}_r)/\partial ({\bf x})$ is the Jacobian
of the transformtion. The reference state has several important 
properties which are straightforward to establish: 1) it depends upon
$z_r$ only, which accounts for the last equality in (\ref{sigma_r}) 
and (\ref{rho_r}); 2) it is in hydrostatic equilibrium at all times,
i.e., $\partial P_r/\partial z_r = -\rho_r g$; 3) the velocity
${\bf u}_r = (dx_r/dt,dy_r/dt,dz_r/dt)$ of the parcels in the reference
state satisfies the mass conservation
equation $D\upsilon_r/Dt = \upsilon_r \nabla_r \cdot {\bf u}_r$.
The main objective of available energetics is to write GPE and IE
as
$$
  GPE = AGPE+GPE_r, \qquad IE = AIE + IE_r
$$
i.e., as the sum of their available components (AGPE and AIE) and
un-available (or reference) components ($GPE_r$ and $IE_r$), and to
derive individual evolution equations for each, as shown below.
%
With regard to $GPE_r$, its definition is
\begin{equation}
    GPE_r = \int_{V_r} \rho_r g z_r \,dV_r = 
   \int_{V} \rho g z_r({\bf x},t)\,dV,
\end{equation}
so that its evolution equation is given by:
\begin{equation}
   \frac{d(GPE_r)}{dt} = \int_{V} \rho g w_r dV = W_r
\end{equation}
by using the definition $Dz_r/Dt=w_r$ and the assumption of
mass conservation of a fluid parcel $D(\rho dV)/Dt=0$. 
With regard to $IE_r$, its definition is
\begin{equation}
    IE_r = \int_{V_r} \rho_r I(\Sigma_r,\upsilon_r)\,dV_r
   = \int_{V} \rho I ( \Sigma, J_r \upsilon) \,dV
\end{equation}
In order to derive an expression for the latter, we use the
differential expression $dIE_r = T_r d\Sigma_r 
- P_r d\upsilon_r = T_r d\Sigma - P_r d\upsilon_r$.
As a result, it can be shown that:
\begin{equation}
   \frac{d(IE_r)}{dt} = D - H - B_r ,
\end{equation}
where
\begin{equation}
   H = \int_{V} \rho \left ( 1 - \frac{T_r}{T} \right )
  \dot{Q} dV, \qquad
    B_r = \int_{V} P_r \frac{D\upsilon_r}{Dt} \rho dV ,
\end{equation}
by using the result that by assumption, $D\Sigma_r/Dt=
D\Sigma/Dt = \dot{Q}/T$. Physically, we interpret $B_r$ and $H$
as the fraction of the internal energy that can be converted into
un-available and available GPE respectively. In the case of $H$,
this is motivated by this term resembling the classical Carnot formula
where the coefficient $(1-T_r/T)$ plays the role of the classical
Carnot efficiency factor, and $\rho \dot{Q}$ the role of the heating source.
In the case of $B_r$, an important result is that it can be rewritten
as
$$
  B_r = \int_{V_r} P_r \frac{D\upsilon_r}{Dt} \rho_r dV_r = 
   - \int_{V_r} {\bf u}_r \cdot \nabla P_r dV_r = W_r
$$
by first rewriting $B_r$ as an integral in the reference space, 
and then using integration by parts and accounting for the abovementioned
three properties of the reference state. Once the evolution equations
for $IE_r$ and $GPE_r$ are known, it is straightforward to derive
evolution equations for AGPE and AIE simply by subtracting those
for GPE and IE minus those for $GPE_r$ and $IE_r$, viz.,
\begin{equation}
   \frac{d(AGPE)}{dt} = W - W_r, \qquad
   \frac{d(AIE)}{dt} = H - B + W_r ,
\end{equation}
taking into account the above result $B_r=W_r$. We now return to the
energetics of the mixing event considered above, by deriving separate
budget equations for AGPE, AIE, $GPE_r$, $IE_r$ as follows:
\begin{equation}
   \Delta AGPE = 0 = \overline{W} - \overline{W}_r , 
   \label{agpe_budget}
\end{equation}
\begin{equation}
   \Delta AIE = 0 = \overline{H} - \overline{B} + \overline{W}_r ,
   \label{aie_budget}
\end{equation}
\begin{equation}
   \Delta GPE_r = \overline{W}_r,
   \label{gper_budget}
\end{equation}
\begin{equation}
    \Delta IE_r = \overline{D} - \overline{H} - \overline{W}_r ,
    \label{ier_budget}
\end{equation}
which in turn implies the following equalities:
\begin{equation}
    \overline{W} = \overline{W}_r = \overline{B}_r = \overline{B}
   -\overline{H}
   \label{central_result} 
\end{equation}
Eqs. (\ref{aie_budget}-\ref{ier_budget}) and Eq.
(\ref{central_result}) are the central result of this paper, for
they prove unambiguously that the ``incompressible'' view of stratified
turbulence
that the ``compressible'' work terms $\overline{B}$ aren $\overline{H}$
are small or even negligible compared to $\overline{W}$ is simply impossible.
Indeed, the result that $\overline{B}-\overline{H} = \overline{W}$ 
in Eq. (\ref{central_result}) imposes
that either or both $\overline{B}$ and $\overline{H}$ be of comparable
magnitude as $\overline{W}$, which is sufficient to refute completely
the validity of the incompressible assumption in presence of irreversible
effects, regardless of the Mach number or of the gas or liquid nature of
the fluid considered. On the other hand, we see no reason to question
the validity and accuracy of the incompressible approximation applied
to {\em adiabatic} motions. In fact, the latter can be applied to the
estimation of $\overline{H}$, since the difference between $T_r$ and
$T$ is entirely due to adiabatic compressibility effects, so that
 $|T-T_r|=0(\Gamma |P-P_r|)$, where $\Gamma = \alpha T/(\rho C_p)$
is the adiabatic lapse rate (with $\alpha$ the thermal expansion,
and $C_p$ the specific heat at constant pressure). In laboratory conditions
at atmospheric pressure, $T-T_r$ is not expected to exceed a few mK for
a fluid such as water or seawater for which $\Gamma = O(10^{-7}\,
K/Pa)$ \cite{Feistel2003}. Although by no means a definitive proof,
this suggests that the above relations can be simplified by neglecting
$\overline{H}$ compared to $\overline{B}$, in which case we expect
$\overline{B} \approx \overline{W}$. Moreover, the new predictions 
for Eqs. (\ref{boussinesq_prediction}) and (\ref{boussinesq_prediction_2})
are given by:
\begin{equation}
     \frac{\Delta KE}{\overline{D}} \approx -1, \qquad
     \frac{\Delta IE}{\overline{D}} \approx (1-\gamma_{mixing}),
     \label{CNSE_prediction}
\end{equation}
\begin{equation}
     \frac{\Delta KE}{\Delta IE} \approx 
     -\frac{1}{1-\gamma_{mixing}} \approx -(1+\gamma_{mixing} 
    + \dots )
    \label{CNSE_prediction_2}
\end{equation}
As stated previously, these new predictions 
 for the ratios $\Delta KE/\overline{D}$ and
$\Delta IE/\overline{D}$ (Eqs. \ref{CNSE_prediction}) differ 
substantially from those resulting from the Boussinesq approximation 
(Eq. \ref{boussinesq_prediction}). Indeed, such ratios would differ
by about $20\%$ for a mixing efficiency $\gamma_{mixing}=0.2$,
the discrepancy between the two models increasing with the measured
mixing efficiency.
On the other hand, both models are found to yield similar predictions
for the ratio $\Delta KE/\Delta IE$ (Eq. \ref{boussinesq_prediction_2}
versus Eq. \ref{CNSE_prediction_2}), so that only the ratios
$\Delta KE/\overline{D}$ and $\Delta IE/\overline{D}$ would be useful
to discriminate between the classical incompressible theory and the
new compressible one presented here.
Whether this can be done with present measurement capabilities is an
issue that is beyond the scope of this paper, but that is would
certainly be of interest to pursue. If not, an alternative would consist in
comparing direct numerical simulations of a turbulent mixing event in
the context of the compressible and incompressible NSE respectively.
Such a project is currently underway.

\par

  In summary, this letter refutes the validity of the incompressible
assumption, which is otherwise well established for adiabatic motions, to describe
fluid flows affected by irreversible diabatic effects.
There does not appear to be a physical
basis, therefore, for regarding dynamics and thermodynamics as being decoupled
in stratified turbulent fluid flows at low Mach numbers, in contrast with the currently
accepted wisdom. The present results are important, because they call for
a complete revisiting of the accepted ideas regarding how stratified turbulence
operates. Some of the several consequences implied by our results are the
basis for \cite{Tailleux2007}, and concern the so-called 
ocean heat engine controversy, and whether turbulence should be regarded
as speeding up the convergence toward thermodynamic equilibrium, with
important implications for turbulent mixing parameterizations in numerical
ocean models. 


\bibliography{pof_arxiv}

\end{document}